\documentstyle[version2,preprint,aps]{revtex}
\tightenlines
\begin{document}

\draft

\begin{title}
Nonuniversal spectral properties of the Luttinger model
\end{title}
\author{K. Sch\"onhammer and V. Meden}
\begin{instit}
Institut f\"ur Theoretische Physik, Universit\"at G\"ottingen,
Bunsenstrasse 9,\\ D-3400 G\"ottingen, Germany
\end{instit}
\receipt{21 December 1992}
\begin{abstract}
The one electron spectral functions for the Luttinger model are
discussed for large but finite systems. The methods presented allow a
simple interpretation of the results. For finite range interactions
interesting nonunivesal spectral features emerge for momenta which
differ from the Fermi points by the order of the inverse interaction
range or more. For a simplified model with interactions only within
the branches of right and left moving electrons analytical
expressions for the spectral function are presented which allows to
perform the thermodynamic limit. As in the general spinless model and
the model including spin for which we present mainly numerical results the
spectral functions do not approach the noninteracting limit for large
momenta. The implication of our results for recent high resolution
photoemission measurements on quasi one-dimensional conductors are
discussed.
\end{abstract}

\section{INTRODUCTION}
\label{sec1}
The experimental study of quasi one-dimensional conductors can
provide a test of the pecularities of correlated electrons in one
dimension \cite{1,2}. In particular high resolution valence
photoemission is a very useful tool, as the measured spectra are
directly related to the one-particle Green's function of the system.
Recent experiments of this type \cite{3} give a strong indication
that the interpretation of spectra requires the inclusion of many
body effects. As theoretical approaches usually concentrate
on the universal behaviour of the spectral functions in the extreme
low energy regime \cite{4} these experiments provide a stimulus to examine
also the {\it nonuniversal behaviour} of the spectral functions for
one-dimensional (1D) correlated electrons.

As first discussed by Tomonaga \cite{5} the problem of 1D electrons
with a {\it long range} interaction simplifies considerably because
it is a good approximation to {\it linearize} the energy dispersion
around the two Fermi points $\pm k_{F}$. In the Luttinger model
\cite{6} an exactly linear dispersion is {\it assumed}. An exact
solution for the Luttinger model was presented by Mattis and Lieb
\cite{7}. The original Tomonaga model and the Luttinger model were
compared by Gutfreund and Schick \cite{8} who showed, that the low
energy physics in both models is the same for long range interaction
with a rather weak restriction on the interaction strength.

The
Luttinger model is often studied with the simplification of a
{\it zero range} interaction \cite{9}. This is sufficient for the
discussion of the low energy singularities of the spectra. As the
interacting ground state of the corresponding model contains holes
deep below the Fermi level, a direct comparison with a system of
nonrelativistic electrons is doubtful, as the linearization of the
energy dispersion is no longer justified for all relevant energies.
We therefore study in this paper the spectral properties of the
Luttinger model for a {\it finite range interaction}.

Even for the spinless case the simplified model with interaction
terms only {\it within} a branch ($g_{4}$-interaction in the
``g-ology''
classification \cite{2}) the nonuniversal behaviour of the spectral
function is very nontrivial for $k$-values which differ from $\pm
k_{F}$ by the order of the inverse interaction range or more. As the
spectral functions of this simplified model can be calculated
analytically this is probably the most simple nontrivial model of
interacting electrons for which a complete explicit calculation of
spectra can be performed. This solution is presented in Sec.
\ref{sec3}
in two different ways: In a direct approach to calculate the many
electron eigenstates which enter the Lehmann representation for the
spectral functions and
using the bosonization of the field operators
\cite{4,10,11}. In both approaches we first calculate the
spectra for systems of finite length $L$ and then perform the
limit $L \rightarrow \infty$.

For the complete spinless Luttinger model including the
$g_{2}$-interaction terms between the branches a recursive numerical
method is presented in Sec. \ref{sec4} to calculate exact spectra for
arbitrarily large systems. This procedure gives more insight into the
interesting nonuniversal features of the $k$ and $\omega$ dependence
of the spectra than the brute force attempt to perform numerically
the double Fourier transform of the Green's functions
$G_{\alpha}^{ \mbox{\tiny ${ > \atop (<) }$ } }(x,t)$ which
themselves have to be calculated involving a numerical integration.
For the $k$-integrated spectral function the direct integration
procedure is compared with the asymptotic results for very large but
finite systems. The approach used for the spinless model is
generalized to the model including spin in Sec. \ref{sec5}. Our results and
the relevance for photoemission spectra are summarized in Sec.
\ref{sec6}. Spectral moments and the momentum distribution are
discussed in Appendices.

The present paper with its discussion of spectral properties at {\it
arbitrary} excitation energies is complementary to our recent work
\cite{11} in which results for the universal low energy part of the
spectra are presented.

\section{Spinless Luttinger model}
\label{sec2}
As discussed in the introduction it is necessary to consider the
Luttinger model with a finite range interaction if one wants the
model to work
as an approximation to describe nonrelativistic electrons not only in
the asymptotic low energy region. The original nonrelativistic
electrons are assumed to interact by a two-body potential $v(x)$,
i.e. the interaction part of the Hamiltonian for a system of
finite length $L$ with periodic boundary conditions reads
\begin{eqnarray}
\label{eqn1}
\hat{V} & = & \frac{1}{2} \int_{0}^{L} dx \int_{0}^{L} dx' v(x-x')
\hat{\rho}(x) \hat{\rho}(x') -\frac{1}{2} v(x=0) \hat{N} \nonumber \\
&=& \frac{1}{2L} \sum_{k \neq 0} \hat{\rho}_{k} \hat{\rho}_{-k}
\tilde{v}(k) + \frac{1}{2L} \hat{N}^{2} \tilde{v}(k=0) - \frac{1}{2}
\hat{N} v(x=0)
\end{eqnarray}
where $\hat{\rho}(x)$ is the operator of the electron density,
$\hat{\rho}_{k}$ with $k=2\pi n/L$, $n \in
{\rm\kern.26em \vrule width.02em height0.5ex depth 0ex
\kern.04em \vrule width.02em height1.47ex depth-1ex \kern-.34em Z}$
its Fourier
components and $\hat{N} \equiv \int \hat{\rho}(x) dx$ the particle
number operator. Here we {\it assume} that the two body potential
$v(x)$ and its Fourier components $\tilde{v}(k)$ are finite at $x=0$
respectively $k=0$.

The transition to the interaction term for the Luttinger model occurs
by writing $\hat{\rho}_{k}$ as
\begin{equation}
\label{eqn2}
\hat{\rho}_{k}=\hat{\rho}_{k,+}+\hat{\rho}_{k,-}
\end{equation}
where the $\hat{\rho}_{k,\alpha}$ with $\alpha=+(-)$ are the Fourier
components of the operators for the densities of right (left) moving
particles. The first term on the rhs. of Eq. (\ref{eqn1}) is generalized to
\cite{2}
\begin{equation}
\label{eqn3}
\hat{V}=\frac{1}{2L} \sum_{k \neq 0, \alpha=\pm} \left( g_{4}(k)
\hat{\rho}_{k,\alpha} \hat{\rho}_{-k,\alpha} + g_{2}(k)
\hat{\rho}_{k,\alpha}  \hat{\rho}_{-k,-\alpha} \right)
\end{equation}
with the original model corresponding to $g_{2}(k) \equiv g_{4}(k)
\equiv \tilde{v}(k)$. The other two terms in Eq. (\ref{eqn1}) are usually
neglected. This is justified for the calculation of the spectral
functions if all frequencies are measured with respect to the
chemical potential $\mu$. For the explicit calculation of $\mu$
these terms are of importance. One comment on the term involving the
interaction $g_{4}(k)$ should be made, as it is often dropped. This
is {\it only} justified for a {\it strictly zero range interaction},
for which the decomposition used in Eq. (\ref{eqn1}) is not allowed.

In the spinless model the $\hat{\rho}_{q,\alpha}$ are given by
\begin{equation}
\label{eqn4}
\hat{\rho}_{q,\alpha} = \sum_{k} \hat{a}_{k,\alpha}^{\dagger}
\hat{a}_{k+q,\alpha}^{}
\end{equation}
where $\hat{a}_{k,\alpha}^{\dagger}$ is the creation operator for a
particle of type $\alpha$ and momentum $k$. With a proper
normalization

\parbox{11.0cm}
{\begin{eqnarray*}
\hat{b}_{q}  \equiv  \left(\frac{2\pi}{|q|L} \right)^{1/2} \times
                             \left\{ \begin{array}{ll}
                             \hat{\rho}_{q,+}  & \mbox{for $q>0$} \\
                             \hat{\rho}_{q,-} &  \mbox{for $q<0$}
                             \end{array}
                             \right.
\end{eqnarray*}}
\hfill
\parbox{2.0cm}
{\begin{equation} \label{eqn5} \end{equation}}

\noindent the density operators obey Bose commutation relations \cite{5,6,7}
\begin{equation}
\label{eqn6}
\left[ \, \hat{b}_{q}^{}\; , \; \hat{b}_{q'}^{\dagger} \, \right]
= \delta_{q,q'} \; \; \; ; \; \; \;
\left[ \, \hat{b}_{q}^{}\; , \; \hat{b}_{q'}^{} \, \right]=0 \; .
\end{equation}
The key to the exact solution of the model lies in the fact, that the
kinetic energy can also be expressed in terms of the Bose operators
\cite{7}
\begin{equation}
\label{eqn7}
\hat{T}=v_{F} \sum_{k,\alpha} \alpha k \hat{a}_{k,\alpha}^{\dagger}
\hat{a}_{k,\alpha}^{} = v_{F} \sum_{q \neq 0} |q| \hat{b}_{q}^{\dagger}
\hat{b}_{q}^{} + c(\hat{N}_{\alpha}) \; .
\end{equation}
The additional term $c(\hat{N}_{\alpha})$ involving the particle number
operators is irrelevant  in the following and will be dropped.
A simple unitary transformation brings the total Hamiltonian
$\hat{H}=\hat{T}+\hat{V}$ into the form \cite{7}
\begin{equation}
\label{eqn8}
\hat{H}=\sum_{q \neq 0} |q| \tilde{v}_{F}(q)
\hat{\alpha}_{q}^{\dagger} \hat{\alpha}_{q}^{}
\end{equation}
where the $\hat{\alpha}_{q}$ are new boson operators and
\begin{equation}
\label{eqn9}
\tilde{v}_{F}(q)= v_{F} \sqrt{\left[1+g_{4}(q)/(2\pi v_{F}) \right]^{2}
-\left[ g_{2}(q)/(2\pi v_{f})\right]^{2}} \; .
\end{equation}
In Eq. (\ref{eqn8}) a constant and terms involving particle number operators
have been dropped in accordance with the discussion of the particle
number terms in Eq. (\ref{eqn1}).

The parts of the one particle Green's function which lead to the
photoemission and inverse photoemission spectra are
\begin{eqnarray}
   \label{eqn11}
   i G_{\alpha}^{<}(x,t) & \equiv & \langle \,
\hat{\psi}_{\alpha}^{\dagger}(0,0)
   \hat{\psi}_{\alpha}^{}(x,t) \, \rangle  \\*
   \label{eqn10}
   i G_{\alpha}^{>}(x,t) & \equiv & \langle \, \hat{\psi}_{\alpha}^{}(x,t)
   \hat{\psi}_{\alpha}^{\dagger}(0,0) \, \rangle
      \; .
\end{eqnarray}
These functions can be calculated exactly e.g. by bosonizing the
field operators $\hat{\psi}_{\alpha}(x,t)$ \cite{4,10}.
For finite systems one obtains \cite{11}
\begin{eqnarray}
   \label{eqn12}
   i G_{\alpha}^{ \mbox{\tiny ${ > \atop (<) }$ } }(x,t) e^{i \mu t}
& = & \frac{1}{L} e^{i\alpha k_{F} x}
  \exp{ \Biggl\{  \frac{2\pi}{L}%
\sum_{q>0} \frac{1}{q} \biggl[ e^{ \mbox{\tiny ${ + \atop (-) }$ } i\alpha qx}
e^{\mbox{\tiny ${ - \atop (+) }$ } i\omega_{q} t}%
 }   \nonumber \\*
&& \hspace{2.5cm}  +2 \, \mbox{s}^{2}(q) \left( \cos{\left(qx\right)}
e^{\mbox{\tiny ${ - \atop (+) }$ } i\omega_{q}%
t}-1\right) \biggr] \Biggr\}
\end{eqnarray}
where $\omega_{q} \equiv |q| \tilde{v}_{F}(q)$ and $\mbox{s}^{2}(q) \equiv
\sinh^{2}{(\Theta_{q})} $ with $\Theta_{q}$ the phase in the unitary
transformation $\hat{\alpha}_{q}=\cosh{(\Theta_{q})} \hat{b}_{q}^{}-
\sinh{(\Theta_{q})} \hat{b}_{-q}^{\dagger}$. From (\ref{eqn12}) one
calculates the relevant spectral function as

\noindent \parbox{11.0cm}
{\begin{eqnarray*}
\rho_{\alpha}^{<}(k,\omega) & \equiv &
\mbox{$\langle \, \phi_{0}^{N} \, | \, $}\hat{a}_{k,\alpha}^{\dagger}
\delta\left(\omega+\left(\hat{H}-E_{0}^{N-1}\right) \right)
\hat{a}_{k,\alpha}^{} \mbox{$|\, \phi_{0}^{N} \, \rangle$}   \\*
& = & \frac{1}{2\pi} \int_{-\infty}^{\infty} dt e^{i\omega t}
\int_{-\infty}^{\infty} dx e^{-ikx} e^{i\mu t} i G_{\alpha}^{<}(x,t)
\; ,
\end{eqnarray*}}
\hfill
\parbox{2.0cm}
{\begin{equation} \label{eqn13} \end{equation}}

\noindent \parbox{11.0cm}
{\begin{eqnarray*}
\rho_{\alpha}^{>}(k,\omega) & \equiv &
\mbox{$\langle \, \phi_{0}^{N} \, | \, $}\hat{a}_{k,\alpha}^{}
\delta\left(\omega-\left(E_{0}^{N+1}-\hat{H}\right) \right)
\hat{a}_{k,\alpha}^{\dagger} \mbox{$|\, \phi_{0}^{N} \, \rangle$}   \\*
& = & \frac{1}{2\pi} \int_{-\infty}^{\infty} dt e^{i\omega t}
\int_{-\infty}^{\infty} dx e^{-ikx} e^{i\mu t} i G_{\alpha}^{>}(x,t)
\; .
\end{eqnarray*}}
\hfill
\parbox{2.0cm}
{\begin{equation} \label{eqn14} \end{equation}}

\noindent The low energy singularities of the spectral functions have been
known for a long time \cite{4,12}. They are obtained by taking the
limit $L\rightarrow \infty$, in which the
$G_{\alpha}^{ \mbox{\tiny ${ > \atop (<) }$ } }(x,t)$ can be calculated
analytically in the large $x$ and $t$ limit. The double Fourier
transform yields expressions for the critical exponents of the
threshold singularities of the spectra.

The aim of our paper is to present results also for the {\it
nonuniversal} frequency range in
$\rho_{\alpha}^{ \mbox{\tiny ${ > \atop (<) }$ } }(k,\omega)$. In
principle this can be done by direct numerical integrations after
performing the limit $L \rightarrow \infty$. As for a finite range
potential the frequencies $\omega_{q}$ have a nontrivial
$q$-dependence, the $G_{\alpha}^{ \mbox{\tiny ${ > \atop (<) }$ } }(x,t)$
have to be calculated involving a numerical integration which has to
be followed by a double numerical Fourier integration. It is
numerically difficult to obtain the sharp spectral features by
this procedure. It also gives little insight into the interpretation
of the calculated spectra. We therefore take a different approach and
calculate the spectra for {\it finite} systems in such a way that the
double Fourier integral can be performed analytically. For the
simplified model with the $g_{4}$-interaction only the limit
$L\rightarrow \infty$ can be directly read off the results for finite
$L$. As this model already shows very interesting nonuniversal
behaviour of the spectra we start with a detailed discussion of this
$g_{4}$-model.

\section{Spectral functions for the spinless \mbox{$g_{4}$}-model}
\label{sec3}
As the Hamiltonian for this special model is a {\it sum} of
Hamiltonians for right and left moving electrons it is sufficient to
consider e.g. the right moving ones only. In the fermion
representation the Hamiltonian reads ($\hat{a}_{n}\equiv
\hat{a}_{k_{n},+}\; ; \; \hat{\rho}_{n} \equiv
\hat{\rho}_{q_{n},+}$)
\begin{equation}
\label{eqn15}
\hat{H}_{+}=\frac{2\pi}{L} \left[ v_{F} \sum_{n} n
\hat{a}_{n}^{\dagger} \hat{a}_{n}^{} + \frac{1}{4\pi} \sum_{n \neq 0}
g_{4}(q_{n}) \hat{\rho}_{n} \hat{\rho}_{-n} \right]
\end{equation}
while the boson representation, apart from an additional term
$c_{+}(\hat{N}_{+})$, is given by ($\hat{b}_{n} \equiv
\hat{b}_{q_{n}}$)
\begin{equation}
\label{eqn16}
\hat{H}_{+}=\frac{2\pi}{L} \left[ v_{F} \sum_{n > 0} n
\hat{b}_{n}^{\dagger} \hat{b}_{n}^{} + \frac{1}{4\pi} \sum_{n > 0} n
g_{4}(q_{n}) \left(\hat{b}_{n}^{\dagger} \hat{b}_{n}^{} +
\hat{b}_{n}^{} \hat{b}_{n}^{\dagger} \right) \right] \; .
\end{equation}
In this representation it is obvious that one can read off the exact
energy eigenvalues {\it without} a canonical transformation. The
eigenstates are {\it identical} to the eigenstates of the {\it
noninteracting system}. Especially the interacting ground state is
given by the Fermi sea $|\, F_{+}(N) \, \rangle$, which has the form
of a Slater determinant. The general eigenstates have the form
\begin{equation}
\label{eqn17}
|\, \left\{ m_{j} \right\}, N \, \rangle = \prod_{j=1}^{\infty}
\left(\frac{1}{m_{j}!}\right)^{1/2} \left(
\hat{b}_{j}^{\dagger}\right)^{m_{j}} |\, F_{+}(N) \, \rangle \; .
\end{equation}
This are {\it linear combinations} of electron-hole pair excited
states which yield the same value of the kinetic energy. It is
instructive to write out the states with low excitation energy to see
the {\it high degeneracy} of states with the same kinetic energy.

The spectral functions for the simplified model can be calculated
directly {\it without} using the bosonization of the field operators.
The spectral weights of the delta peaks e.g. in
$\rho_{+}^{>}(k_{n},\omega)$ are given by $\left| \, \langle \,
\left\{ m_{j} \right\},N+1 \, | \, \hat{a}_{n}^{\dagger} \,
| \, F_{+}(N) \, \rangle \, \right|^{2}$ and can be calculated using
$\hat{a}_{n}^{\dagger} | \, F_{+}(N) \, \rangle=\hat{a}_{n}^{\dagger}
\hat{a}_{n_{F}+1}^{} | \, F_{+}(N+1) \, \rangle$, where $n_{F}=L
k_{F}(N)/(2\pi)$, and
\begin{equation}
\label{eqn18}
\hat{b}_{n}^{\dagger}=\left(\frac{1}{n}\right)^{1/2} \sum_{m}
\hat{a}_{m+n}^{\dagger} \hat{a}_{m}^{}
\end{equation}
for $n\geq 1$. For $n=n_{F}+1$ and $n=n_{F}+2$ the states
$\hat{a}_{n}^{\dagger} | \, F_{+}(N) \, \rangle$ are eigenstates:
$\hat{a}_{n_{F}+1}^{\dagger} | \, F_{+}(N) \, \rangle= | \, F_{+}(N+1) \,
\rangle$,
$\hat{a}_{n_{F}+2}^{\dagger} | \, F_{+}(N) \, \rangle=
\hat{b}_{1}^{\dagger}| \, F_{+}(N+1) \,
\rangle$. For $n=n_{F}+1+\tilde{n}$ with $\tilde{n}\geq 1$ the state
$\hat{a}_{n}^{\dagger} | \, F_{+}(N) \, \rangle =
\hat{a}_{n_{F}+1+\tilde{n}}^{\dagger} \hat{a}_{n_{F}+1}^{}
| \, F_{+}(N+1) \, \rangle$ has overlap to the states
$\hat{b}_{\tilde{n}}^{\dagger} | \, F_{+}(N+1) \, \rangle $,
$\hat{b}_{1}^{\dagger} \hat{b}_{\tilde{n}-1}^{\dagger}
| \, F_{+}(N+1) \, \rangle$,
$\hat{b}_{2}^{\dagger} \hat{b}_{\tilde{n}-2}^{\dagger}
| \, F_{+}(N+1) \, \rangle$,
$\left(1/2!\right)^{1/2} \left(\hat{b}_{1}^{\dagger}\right)^{2}
\hat{b}_{\tilde{n}-2}^{\dagger} | \, F_{+}(N+1) \, \rangle$ etc.,
i.e. to states $|\, \left\{ m_{j} \right\}, N+1 \, \rangle$ with
$\sum_{j \geq 1} j m_{j}=\tilde{n}$. For large $\tilde{n}$ these
eigenstates are rather complicated linear combinations of
electron-hole pair excited states, but the expansion coefficient of
the component
$\hat{a}_{n_{F}+1+\tilde{n}}^{\dagger} \hat{a}_{n_{F}+1}^{}
| \, F_{+}(N+1) \, \rangle$ is simple.
The square of the overlap is
given by
\begin{equation}
\label{eqn19}
\left| \, \langle \,
\left\{ m_{j} \right\},N+1 \, | \, \hat{a}_{n_{F}+1+\tilde{n}}^{\dagger} \,
| \, F_{+}(N) \, \rangle \, \right|^{2} = \prod_{j \geq 1}
\frac{1}{m_{j}!} \left( \frac{1}{j}\right)^{m_{j}} \equiv
A( \{ m_{j} \} ) \; .
\end{equation}
The factor $1/m_{j}!$ is due to the corresponding factor in
Eq. (\ref{eqn17}) while the factor $(1/j)^{m_{j}}$ comes from the prefactor
on the rhs. of Eq. (\ref{eqn18}). This yields for the spectral function
\begin{equation}
\label{eqn20}
\rho^{>}(k_{n_{F}+1+\tilde{n}} , \omega) = \sum_{\{ m_{j} \} }
\delta_{\sum_{j} j m_{j},\tilde{n}} A(\{ m_{j} \})
\delta\left( \omega-\sum_{j} m_{j} \omega_{j} \right)
\end{equation}
where $\omega_{j}=(2\pi /L)(v_{F}+g_{4}(q_{j})/(2\pi))$.
Alternatively this result can be obtained using Eq. (\ref{eqn12}) for
$\mbox{s}^{2}(q)\equiv 0$ by formally expanding the exponential
function and performing the double Fourier integral in Eq.
(\ref{eqn14})
analytically. We will discuss this procedure in more detail later.

The calculation of $\rho^{>}(k_{n},\omega)$ is therefore reduced to
the combinational problem to find all decompositions
\begin{equation}
\label{eqn21}
m_{1}+2m_{2}+3m_{3}+\ldots +\tilde{n}m_{\tilde{n}}=\tilde{n}
\end{equation}
with $m_{j} \in {\rm I\kern-.23em N}_{0}$. The solution can be easily produced
on a
computer, but for large $\tilde{n}$ the number of decompositions
increases exponentially. For the special $q$-dependence of $g_{4}$ used in the
following
\begin{equation}
\label{eqn22}
g_{4}(k)=g_{4} \Theta(k_{c}^{2}-k^{2})
\end{equation}
where $r_{c}=1/k_{c}$ is the effective range of the interaction,
we will therefore also use a different technique. The $k$-dependent
Fermi velocity $\tilde{v}_{F}(k)$ takes only two different values with this
assumption

\parbox{11.0cm}
{\begin{eqnarray*}
\tilde{v}_{F}(k)=\left\{ \begin{array}{ll}
                             \tilde{v}_{F}=v_{F}\left(
                             1+g_{4}/\left(2\pi v_{F} \right) \right)
                             & \mbox{for $0<k\leq k_{c}$} \\
                             v_{F} &  \mbox{for $k> k_{c}$}  \; .
                             \end{array}
                             \right.
\end{eqnarray*}}
\hfill
\parbox{2.0cm}
{\begin{equation} \label{eqn23} \end{equation}}

\noindent For a finite system this means that $g_{4}(k_{n})=g_{4}$ for $1 \leq
n \leq
n_{c}$ where $n_{c}=Lk_{c}/(2\pi)$ and $g_{4}(k_{n})=0$ for
$n>n_{c}$.

 From Eq. (\ref{eqn20}) it follows that $\rho^{>}(k_{n},\omega)$ is
{\it trivial for} $0<k_{n}-k_{F}<k_{c}$, as $m_{j}$ can only be
different from zero for $j\leq \tilde{n}$ and {\it all} corresponding
$\omega_{j}$ are given by $(2\pi/L)\tilde{v}_{F}$. This yields
($\tilde{k}=k-k_{F}$)
\begin{equation}
\label{eqn24}
\rho^{>}(k_{F}+\tilde{k},\omega)=\delta(\omega-\tilde{v}_{F}
\tilde{k}) \;\; \mbox{for}\;\; 0<\tilde{k}<k_{c}
\end{equation}
i.e. Fermi liquid like behaviour. This is due to the special choice
(\ref{eqn22}) for $g_{4}(k)$. For $\tilde{k}>k_{c}$ the spectral
functions are nontrivial. One has to distinguish the intervals
$mk_{c}<\tilde{k}<(m+1)k_{c}$. We discuss in the following small
values of $m$ and the limit $m \rightarrow \infty$. For $m=1$ it is
still very simple to argue in terms of the decompositions in
Eq. (\ref{eqn21}). For $n_{c}<\tilde{n}<2n_{c}$ one can have at most
{\it one} nonzero $m_{l}$ ($m_{l}=1$) for $n_{c}<l\leq \tilde{n}$.
The remaining ``momentum'' $\tilde{n}-l$ can be decomposed into momenta
$j$ which are smaller than $n_{c}$. Therefore the energy for all
these decompositions is given by
$(2\pi/L)\left[ lv_{F}+(\tilde{n}-l) \tilde{v}_{F} \right]
=(2\pi/L)\left[ \tilde{n} v_{F}+(\tilde{n}-l) (\tilde{v}_{F}-v_{F})
\right]$. The correponding weight is $1/l$. The remaining weight lies
in a delta peak at $(2\pi/L) \tilde{n} \tilde{v}_{F}$, which
corresponds to the decompositions of $\tilde{n}$ with nonzero $m_{j}$
only for $j \leq n_{c}$. The limit $L\rightarrow \infty$ can easily
be read off and one obtains
\begin{eqnarray}
\label{eqn25}
\rho^{>}(k_{F}+\tilde{k},\omega) & = &\frac{\Theta(\omega-v_{F}\tilde{k})%
\Theta(v_{F}\tilde{k}+(\tilde{v}_{F}-v_{F})(\tilde{k}-k_{c})-\omega)}%
{\tilde{v}_{F} \tilde{k} -\omega} \nonumber \\
&& + \left[ 1-
\ln{\left(\tilde{k}/k_{c}\right)} \right] \delta(\omega-\tilde{v}_{F}
\tilde{k}) \; .
\end{eqnarray}
The weight $z(\tilde{k})$ of the delta peak decreases continously from $1$ to
$1-\ln{2} \approx 0.307$ when $\tilde{k}$ increases from $k_{c}$ to
$2k_{c}$. The shape of the spectral weight is
shown in Fig. 1(b). If one further increases $\tilde{k}$ the calculation
of the spectrum using Eq. (\ref{eqn20}) becomes more and more tedious.
We will therefore analyse the spectra by another method.

In the limit
$\tilde{k} \rightarrow \infty$ one might expect that the spectral
function $\rho^{>}(k_{F}+\tilde{k},\omega)$ reduces to a delta
function as for noninteracting electrons. That this is {\it not} the
case can be seen quite generally by calculating the first and second
moment of $\rho^{>}(k_{F}+\tilde{k},\omega)$. As shown in Appendix A
the result for
$\Delta_{\tilde{k}}^{2}=\mu_{2}^{>}(k_{F}+\tilde{k})-
(\mu_{1}^{>}(k_{F}+\tilde{k}))^{2}$ for the special model of
Eq. (\ref{eqn22}) for $\tilde{k}>2k_{c}$ is given in the limit $L\rightarrow
\infty$ by
\begin{equation}
\label{eqn26}
\Delta_{\tilde{k}}^{2}=\frac{1}{2}
k_{c}^{2}(\tilde{v}_{F}-v_{F})^{2}=
\frac{1}{2}\left(\mu_{1}^{>}(k_{F}+\tilde{k})-v_{F} \tilde{k}
\right)^{2}
\end{equation}
i.e. the effective width of the spectrum is {\it independent} of
$\tilde{k}$ as soon as $\tilde{k}$ is larger than $2k_{c}$.

We now present a method to calculate the spectra using
Eq. (\ref{eqn12}). As in the following we measure energies with respect to
$\mu$ and
momenta with respect to $k_{F}$ we have
$iG^{>}(x,t)=\exp{[F(x,t)]}/L$, where for $g_{4}(k)$ given by
Eq. (\ref{eqn22})
\begin{equation}
\label{eqn27}
F(x,t)=\sum_{n=1}^{n_{c}} \frac{1}{n} \exp{\left( i\frac{2\pi}{L} n%
\left[x-\tilde{v}_{F}t \right] \right)} + \sum_{n=n_{c}+1}^{\infty} \frac{1}{n}
\exp{\left(i\frac{2\pi}{L} n \left[x-v_{F}t \right] \right)} \; ,
\end{equation}
as $\mbox{s}^{2}(q) \equiv 0$ for $g_{2}(k) \equiv 0$. The second sum
on the rhs. of Eq. (\ref{eqn27}) is not convergent as it stands. There
are two obvious methods to overcome this difficulty. One can add a
factor $\exp{(-0n)}$ or restrict the sum to $n \leq M$, where
$M(2\pi/L)$ is much larger than the momenta one is interested in.
Both procedures give the {\it same} results for the spectra. Using
the first method we can write Eq. (\ref{eqn27}) as
\begin{eqnarray}
\label{eqn28}
F(x,t) & =&
-\ln{\left(1-\exp{\left(i\frac{2\pi}{L}\left[x-v_{F}t+i0\right]%
\right)}\right)} \nonumber \\
&& + \sum_{n=1}^{n_{c}} \frac{1}{n}
\left( \exp{\left(i\frac{2\pi}{L} n \left[x-\tilde{v}_{F}t \right] \right)}
-\exp{\left(i\frac{2\pi}{L} n \left[x-v_{F}t \right] \right)} \right)
\, .
\end{eqnarray}
Unfortunatly the finite sums cannot be summed in closed form. We therefore
write
\begin{eqnarray}
iG^{>}(x,t) & = &
\frac{1/L}{1-\exp{\left(i\frac{2\pi}{L}\left[x-v_{F}t+i0\right]\right)}}
\nonumber  \\*
&& \times \prod_{m=1}^{n_{c}}  \left(\sum_{j=1}^{\infty} \frac{(-1/m)^{j}}{j!}
\exp{\left(i\frac{2\pi}{L}
mj\left[x-v_{F}t\right]\right)} \right)  \nonumber \\*
&& \hspace{1.2cm} \times \left(\sum_{l=1}^{\infty} \frac{(1/m)^{l}}{l!}
\exp{\left(i\frac{2\pi}{L}
ml\left[x-\tilde{v}_{F}t\right]\right)} \right) \nonumber
\end{eqnarray}
\begin{eqnarray}
\label{eqn29}
 =  \frac{1}{L} \sum_{m=0}^{\infty} a_{m}^{(n_{c})}
\exp{\left(i\frac{2\pi}{L} m \left[x-v_{F}t\right] \right)}
\sum_{l=0}^{\infty} b_{l}^{(n_{c})}
\exp{\left(i\frac{2\pi}{L} l \left[x-\tilde{v}_{F}t\right] \right)}
\end{eqnarray}
where the $a_{m}^{(n_{c})}$ and $b_{l}^{(n_{c})}$ can be determined
iteratively for $m \geq 1$, $l \in {\rm I\kern-.23em N}_{0}$ and $i=0, \ldots
,m-1$

\parbox{11.0cm}
{\begin{eqnarray*}
a_{lm+i}^{(m+1)} & = & \sum_{j=0}^{l} \frac{(-1/m)^{j}}{j!} a_{m(l-j)+i}^{(m)}
\\
b_{lm+i}^{(m+1)} & = & \sum_{j=0}^{l} \frac{(1/m)^{j}}{j!} b_{m(l-j)+i}^{(m)}
\; .
\end{eqnarray*}}
\hfill
\parbox{2.0cm}
{\begin{equation} \label{eqn30} \end{equation}}

\noindent The starting values are $b_{m}^{(1)}=1/m!$ and
\begin{equation}
\label{eqn31}
a_{m}^{(1)}=\sum_{j=0}^{m} (-1)^{j}/j! \; .
\end{equation}
Using Eq. (\ref{eqn23}) the double Fourier transform can be trivially
performed ($k_{F}\equiv 0$)
\begin{equation}
\label{eqn32}
\rho^{>}(k_{n},\omega)=\sum_{l=0}^{n} a_{n-l}^{(n_{c})}
b_{l}^{(n_{c})} \delta \left(\omega-\frac{2\pi}{L}  \left[ \left(
n-l \right) v_{F} +l \tilde{v}_{F} \right] \right) \; .
\end{equation}
Compared to Eq. (\ref{eqn20}) this representation avoids the
combinatorial problem of Eq. (\ref{eqn21}) and can be used
numerically for much larger values of $n_{c}$ than Eq. (\ref{eqn20}).
Some important information about the $a_{m}^{(n_{c})}$ and
$b_{m}^{(n_{c})}$ can be obtained {\it analytically}. For the case of
the $a_{i}^{(n_{c})}$ it is useful to go back to Eq. (\ref{eqn27}), where
one can directly read off $a_{0}^{(n_{c})}=1$, $a_{i}^{(n_{c})}=0$
for $i=1,\ldots ,n_{c}$, $a_{i}^{(n_{c})}=1/i$ for
$i=n_{c}+1, \ldots,2n_{c}+1$ etc.. The limit $m\rightarrow \infty$ of
$a_{m}^{(n_{c})}$ for fixed $n_{c}$ follows from Eqs. (\ref{eqn30})
and (\ref{eqn31})
\begin{equation}
\label{eqn33}
a_{m}^{(n_{c})} \rightarrow \prod_{n=1}^{n_{c}}
\left(\frac{1}{e}\right)^{1/n} \; .
\end{equation}
In the large $n_{c}$ limit one therefore obtains
$a_{m}^{(n_{c})}\rightarrow e^{-C}/n_{c} \approx 0.56/n_{c}$ where $C$
is Euler's constant. The behaviour of the $b_{m}^{(n_{c})}$ for $m$
values of the order of $n_{c}$ follows from Eq. (\ref{eqn27}) if one
writes the first term on the rhs. as a difference of a logarithm
similar to the first term on the rhs. of Eq. (\ref{eqn28}) and the
sum running from $n_{c}+1$ to infinity. This yields
$b_{i}^{(n_{c})}=1$, for $i=1,\ldots ,n_{c}$
\begin{equation}
\label{eqn34}
b_{i}^{(n_{c})}=1-\left(\frac{1}{n_{c}+1}+\frac{1}{n_{c}+2}+\ldots
+\frac{1}{i} \right)
\end{equation}
for $n_{c}+1 \leq i < 2n_{c}$, etc..

In this approach it is very simple to obtain the {\it exact
analytical result} for the spectrum in the {\it large momentum limit}
$k \gg k_{c}$
\begin{equation}
\label{eqn35}
\rho^{>}(k_{n},\omega) \rightarrow \prod_{m=1}^{n_{c}} \left(
\frac{1}{e}\right)^{1/m} \sum_{l=0}^{\infty} b_{l}^{(n_{c})}
\delta \left(\omega - \frac{2\pi}{L}\left[nv_{F}+l\left(
\tilde{v}_{F}-v_{F}\right) \right] \right) \; .
\end{equation}
We discuss this for the special case $n_{c}=1$ and $n_{c}\rightarrow
\infty$ only. For $n_{c}=1$ we have $b_{l}^{(1)}=1/l!$ and the
spectrum is a {\it Poisson distribution} with a unit strength
parameter. In the thermodynamic limit $L\rightarrow \infty$,
$n_{c}\rightarrow \infty$ with $k_{c}=n_{c}(2\pi/L)=\mbox{const}$
the spectrum is given by ($\delta v_{F} \equiv \tilde{v}_{F}-v_{F}$)

\parbox{11.0cm}
{\begin{eqnarray*}
\rho^{>}(k,\omega)=\frac{e^{-C}}{\delta v_{F} k_{c}} \times
                             \left\{ \begin{array}{ll}
                             1   & \mbox{for $0<\omega-v_{F} k<\delta
                             v_{F} k_{c} $} \\
                             1-\ln{\left(\frac{\omega-v_{F}k}{\delta%
                              v_{F} k_{c}} \right)}
                              &  \mbox{for $\delta v_{F}k_{c}
                              <\omega-v_{F} k<2\delta v_{F} k_{c}$} \; ,
                             \end{array}
                             \right.
\end{eqnarray*}}
\hfill
\parbox{2.0cm}
{\begin{equation} \label{eqn36} \end{equation}}

\noindent etc.. This behaviour is shown in Fig. 1(d). Fig. 1 summarizes our
results
for the $g_{4}$-model for $\tilde{v}_{F}=1.2v_{F}$, i.e. a repulsive
interaction. In Fig. 1(a) all spectral weight lies in a delta peak at
$\omega=\tilde{v}_{F}k$ ($z=1$). For $k>k_{c}$ there continues to be a delta
peak at $\omega=\tilde{v}_{F}k$, but its weight decreases rapidly
with increasing momentum as shown in Figs. 1(b) and 1(c). The
additional weight lies in the continous part of the spectrum as discussed
analytically in Eqs. (\ref{eqn25}) and (\ref{eqn36}). In the limit $k
\gg k_{c}$ one does not recover the limit of noninteracting
electrons. The {\it shape} of the spectrum becomes {\it independent} of
the interaction strength as a function of the scaled variable
$(\omega-v_{F} k_{F})/(k_{c}\delta v_{F})$. A calculation of the
spectrum using perturbation theory for the selfenergy to low order
(e.g. second order) completely fails to give the correct shape of the
spectrum.

Another quantity of interest is the total spectral density per unit
length
\begin{eqnarray}
\label{eqn37}
\rho^{>}(\omega) & = & \int_{0}^{\infty} \frac{dk}{2\pi}
\rho^{>}(k,\omega)   \\
& = & \frac{1}{2\pi}\int_{-\infty}^{\infty} dt G^{>}(0,t) \nonumber
\; .
\end{eqnarray}
For the special $g_{4}$-interaction (\ref{eqn22}) the function
$\rho^{>}(k,\omega)$ has been given analytically in Eqs.
(\ref{eqn24}) and (\ref{eqn25}). The $k$-integration can be simply
performed and yields

\parbox{11.0cm}
{\begin{eqnarray*}
\rho^{>}(\omega)=\frac{1}{2\pi \tilde{v}_{F}} \times
                             \left\{ \begin{array}{ll}
                             1   & \mbox{for $0 \leq \omega<v_{F} k_{c}$} \\
                             1+\ln{\left(\frac{\omega}{v_{F} k_{c}} \right)}
                              &  \mbox{for $v_{F}k_{c}<\omega<
                              \tilde{v}_{F} k_{c}$} \\
                             1+\ln{\left(\frac{\tilde{v}_{F}}{v_{F}} \right)}
                              &  \mbox{for $\tilde{v}_{F}k_{c}<\omega<
                              2v_{F} k_{c}$}  \; ,
                             \end{array}
                             \right.
\end{eqnarray*}}
\hfill
\parbox{2.0cm}
{\begin{equation} \label{eqn38} \end{equation}}

\noindent etc.. Alternatively the function $F(0,t)$ in Eq. (\ref{eqn27}) can be
calculated numerically in the limit $L\rightarrow \infty$ and
$G^{>}(0,t)$ is Fourier transformed numerically. The results in both
approaches agree and are shown in Fig. 2. The latter method can be
performed for {\it arbitrary} interactions $g_{4}(k)$. Results for
$g_{4}(k)=g_{4} \exp{(-|k|/k_{c})}$ are also shown in Fig. 2 for a
positive and a negative value of $g_{4}$. For the case of a repulsive
interaction there is a {\it depletion} of the total spectral weight
at low frequencies. Due to a sum rule discussed in \cite{13} the
missing weight has to show up in another frequency range. For a model
without an upper cut-off in momentum space the missing weight is
pushed to {\it infinity}. If a cut-off is included the missing weight
appears at the upper end of the spectrum, as the spectral weight in
Fig. 1(d) has to be compared to a delta peak at $v_{F}k_{max}$ for
noninteracting electrons.

In this section the simple $g_{4}$-model has been discussed quite at
length, as the generalization to the complete model including $g_{2}$
terms has to be performed mainly numerically.


\section{Spectral functions for the general spinless model}
\label{sec4}
The method to calculate $G^{>}(x,t)$ presented in Eq. (\ref{eqn29})
is generalized to the full spinless model introduced in Sec.
\ref{sec2}. We specialize to interactions of the type described in
(\ref{eqn22}), i.e. we also assume $g_{2}(k)=g_{2}
\Theta(k_{c}^{2}-k^{2})$. Then using Eq. (\ref{eqn12}) the expression
for $F(x,t)$ in (\ref{eqn27}) is generalized to ($s^{2}=s^{2}(q_{n})$
for $1 \leq n \leq n_{c}$)
\begin{eqnarray}
\label{eqn39}
F(x,t) & = & \sum_{n=1}^{n_{c}} \frac{1}{n} \left[ \left(1+s^{2}\right)
\exp{\left( i\frac{2\pi}{L} n%
\left[x-\tilde{v}_{F}t \right] \right)}+ s^{2}
\exp{\left(-i\frac{2\pi}{L} n%
\left[x+\tilde{v}_{F}t \right] \right)} \right. \nonumber \\*
&& \left. \hspace{-1.3cm} -2s^{2}-
\exp{\left( i\frac{2\pi}{L} n%
\left[x-{v}_{F}t \right] \right)} \right]
-\ln{\left[1-\exp{\left(i\frac{2\pi}{L} n \left[x-v_{F}t+i0 \right]
\right)}\right]}   \, .
\end{eqnarray}
Again $iG^{>}(x,t)=\exp{[F(x,t)]}/L$ is written as a product of power
series
\begin{eqnarray}
\label{eqn40}
iG^{>}(x,t) & = & \frac{1}{L} \left(\sum_{m=0}^{\infty} a_{m}^{(n_{c})}
\exp{\left(i\frac{2\pi}{L} m \left[x-v_{F}t\right] \right)} \right)
\nonumber \\*
&& \times \left( \sum_{l=0}^{\infty} b_{l}^{(n_{c})}
\exp{\left(i\frac{2\pi}{L} l \left[x-\tilde{v}_{F}t\right] \right)}
\right) \nonumber \\*
&& \times
\left( \sum_{r=0}^{\infty} c_{r}^{(n_{c})}
\exp{\left(-i\frac{2\pi}{L} r \left[x+\tilde{v}_{F}t\right] \right)}
\right) \exp{\left(-2s^{2}\sum_{n=1}^{n_{c}}\frac{1}{n}\right)} \, ,
\end{eqnarray}
where the expansion coefficients are determined iteratively as in
Sec. \ref{sec3}. For $m \geq 1$, $l \in {\rm I\kern-.23em N}_{0}$ and $i=0,
\ldots ,m-1$
one obtains

\parbox{11.0cm}
{\begin{eqnarray*}
a_{lm+i}^{(m+1)} & = & \sum_{j=0}^{l} \frac{(-1/m)^{j}}{j!} a_{m(l-j)+i}^{(m)}
\\
b_{lm+i}^{(m+1)} & = & \sum_{j=0}^{l} \frac{\left(
\left[1+s^{2}\right] /m \right)^{j}}{j!} b_{m(l-j)+i}^{(m)} \\
c_{lm+i}^{(m+1)} & = & \sum_{j=0}^{l} \frac{\left(
s^{2} /m \right)^{j}}{j!} c_{m(l-j)+i}^{(m)} \; .
\end{eqnarray*}}
\hfill
\parbox{2.0cm}
{\begin{equation} \label{eqn41} \end{equation}}

\noindent The starting values are $b_{m}^{(1)}=(1+s^{2})^{m}/m!$,
$c_{m}^{(1)}=s^{2m}/m!$ and $a_{m}^{(1)}$ is given by Eq.
(\ref{eqn31}). Using Eq. (\ref{eqn40}) the double Fourier transform
can be simply performed ($A=\exp{\left(\sum_{n=1}^{n_{c}} 1/n
\right)}$)
\begin{eqnarray}
\label{eqn42}
\rho^{>}(k_{n},\omega) & = & A^{-2s^{2}}  \sum_{r=0}^{\infty}
\sum_{j=0}^{n+r}
c_{r}^{(n_{c})} a_{n+r-j}^{(n_{c})}
b_{j}^{(n_{c})} \nonumber \\*
&& \times \delta \left(\omega-\frac{2\pi}{L}  \left[ \left(
n+r-j \right) v_{F} +\left( r+j \right) \tilde{v}_{F} \right] \right) \; .
\end{eqnarray}
If we write $(n+r-j) v_{F}+ (r+j)
\tilde{v}_{F}=nv_{F}+j(\tilde{v}_{F}-v_{F})+r ( \tilde{v}_{F}-v_{F})$ it
is obvious that $\rho^{>}(k_{n},\omega) \equiv 0$ for $\omega < k_{n}
v_{F}$ is guaranteed only for $\tilde{v}_{F}-v_{F}>0$, i.e. for {\it
repulsive} interactions if $g_{4}=g_{2}$.

The coefficients $a_{m}^{(n_{c})}$ are the same as in Sec.
\ref{sec3}. The behaviour of the $b_{m}^{(n_{c})}$ and
$c_{m}^{(n_{c})}$ for $m \leq n_{c}$ follow from the identity
\begin{eqnarray}
\label{eqn43}
\exp{\left( \gamma \sum_{n=1}^{n_{c}}\frac{1}{n} z^{n} \right) } & = &
\left( 1-z \right)^{-\gamma}
\exp{\left(-\gamma \sum_{n=n_{c}+1}^{\infty}\frac{1}{n} z^{n} \right) }
\nonumber \\
& = & \left[ 1+ \sum_{m=1}^{\infty} \left( \prod_{j=1}^{m} \left[
1+\frac{\gamma-1}{j} \right] \right) z^{m} \right] \nonumber \\ && \times
\exp{\left(-\gamma \sum_{n=n_{c}+1}^{\infty}\frac{1}{n} z^{n} \right) }
\; .
\end{eqnarray}
For $1 \leq m \leq n_{c}$ the expansion coefficients are therefore
given by the first factor on the rhs. of Eq. (\ref{eqn43}), i.e.

\parbox{11.0cm}
{\begin{eqnarray*}
b_{m}^{(n_{c})} & = & \prod_{j=1}^{m} \left( 1+\frac{s^{2}}{j}\right)
\stackrel{n_{c} \geq m \gg 1}{\longrightarrow} \mbox{const} \, \times m^{s^{2}}
\\
c_{m}^{(n_{c})} & = & \prod_{j=1}^{m} \left( 1+\frac{s^{2}-1}{j}\right)
\stackrel{n_{c} \geq m \gg 1}{\longrightarrow} \mbox{const} \, \times
m^{s^{2}-1}
\end{eqnarray*}}
\hfill
\parbox{2.0cm}
{\begin{equation} \label{eqn44} \end{equation}}

\noindent This power law behaviour of the coefficients $b_{m}^{(n_{c})}$ and
$c_{n}^{(n_{c})}$ is responsible for the low energy power law
singularities of the spectral functions and the power law behaviour
of the momentum distribution $n(k)$ discussed in Appendix
\ref{apenb}. To demonstrate this we consider momenta $1 \ll n \ll
n_{c}$ and frequencies of the order $\tilde{v}_{F}(2\pi n/L)$ in Eq.
(\ref{eqn42}). As discussed before Eq. (\ref{eqn33}),
$a_{m}^{(n_{c})}= \delta_{m,0}$ for $m \leq n_{c}$. Therefore the
spectral function simplifies in this regime to
\begin{equation}
\label{eqn45}
\rho^{>}(k_{n},\omega)=\sum_{r \geq 0} c_{r}^{(n_{c})}
b_{n+r}^{(n_{c})} \delta \left( \omega-\frac{2\pi}{L}
\left[n+2r\right] \tilde{v}_{F} \right) \; ,
\end{equation}
i.e. the spectrum consists of delta peaks at $\omega=\tilde{v}_{F}
k_{n}+\tilde{v}_{F}(4\pi r /L)$. For $1 \ll r+n \ll n_{c}$ we can use
the asymptotic form of the coefficients in Eq. (\ref{eqn44}) at
$r=(\omega -\tilde{v}_{F} k_{n})/(4\pi \tilde{v}_{F}/L)$ and
$n+r=(\omega +\tilde{v}_{F} k_{n})/(4\pi \tilde{v}_{F}/L)$
to obtain the weights of the peaks. In the limit $L\rightarrow \infty$
this yields
\begin{equation}
\label{eqn46}
\rho^{>}(k_{n},\omega) \sim \Theta(\omega-\tilde{v}_{F} k)
(\omega-\tilde{v}_{F} k)^{s^{2}-1} (\omega+\tilde{v}_{F} k)^{s^{2}}
\; ,
\end{equation}
i.e. the well known asymptotic behaviour for $k \ll k_{c}$ and
$\omega - \tilde{v}_{F} k \ll \tilde{v}_{F} k_{c}$ \cite{4,11}.

In the opposite limit $k \gg k_{c}$ the coefficients
$a_{n+r-j}^{(n_{c})}$ in Eq. (\ref{eqn42}) can be replaced by the
constant introduced in Eq. (\ref{eqn33}). Near the threshold at
$\omega = v_{F} k_{n}$ Eq. (\ref{eqn44}) can be used again for the
$c_{r}^{(n_{c})}$ and $b_{j}^{(n_{c})}$. Performing the integrations
in the limit $L \rightarrow \infty$ the finite step at threshold in
Eq. (\ref{eqn36}) is replaced by a power law behaviour $(\omega -v_{F}
k)^{2s^{2}}$. For arbitrary values of $k_{n}$ and $\omega$ the
spectral function $\rho^{>}(k_{n},\omega)$ has to be calculated
numerically. Results for the same $k$ values as used for the
$g_{4}$-model in Sec. \ref{sec3} and $g_{2}(k) \equiv g_{4}(k)$
are shown in Figs. 3(a)-(d). The figures shows that the delta peak at
$\omega=\tilde{v}_{F} k$ is changed into a power law behaviour and
one could analytically show that the critical exponent for $\omega
\searrow \tilde{v}_{F} k$ is given by $s^{2}-1$ as in Eq.
(\ref{eqn46}). The shape of the continous part of the spectra of the
$g_{4}$-model is modified quite considerably for the parameters used
($s^{2}=1/8$ corresponding to $\tilde{v}_{F}=2 v_{F}$). In contrast
to the $g_{4}$-model $\rho^{>}(k,\omega) \neq 0$ for $k<0$ and small
$|k|$ \cite{11}, but the spectral weight for negative $k$ decreases
rapidly for increasing $|k|$.

In Fig. 4 we show the integrated spectral weight for $g_{4}(k) \equiv
g_{2}(k)$ for the step model (\ref{eqn22}) as well as the exponential
model $g_{4}(k) \equiv g_{2}(k)=g \exp{(-|k|/k_{c})}$ used also in
Fig. 2. At low frequencies the power law behaviour proportional to
$\omega^{2s^{2}}$ leads to a suppression of spectral weight \cite{14}.
For the case of an {\it attractive} interaction this leads to a peak
in $\rho^{>}(\omega)$.

\section{Generalizations for the model including spin}
\label{sec5}
For the Luttinger model including spin the discussion in Sec.
\ref{sec2} applies up to Eq. (\ref{eqn4}). If the density operators
$\hat{\rho}_{q,\alpha}$ are decomposed into a sum of particle-hole
operators an additional spin summation occurs. It is then useful to
define charge and spin operators \cite{15}

\parbox{11.0cm}
{\begin{eqnarray*}
\hat{\rho}_{q,\alpha} & \equiv & \hat{\rho}_{q,\alpha,\uparrow} +
\hat{\rho}_{q,\alpha,\downarrow} \\*
\hat{\sigma}_{q,\alpha} & \equiv & \hat{\rho}_{q,\alpha,\uparrow} -
\hat{\rho}_{q,\alpha,\downarrow} \; .
\end{eqnarray*}}
\hfill
\parbox{2.0cm}
{\begin{equation} \label{eqn47} \end{equation}}

\noindent With a normalization which differs by a factor of $\sqrt{2}$ the
analogous definition to Eq. (\ref{eqn5}) reads

\parbox{11.0cm}
{\begin{eqnarray*}
\hat{b}_{q,c}  \equiv  \left(\frac{\pi}{|q|L} \right)^{1/2} \times
                             \left\{ \begin{array}{ll}
                             \hat{\rho}_{q,+}  & \mbox{for $q>0$} \\*
                             \hat{\rho}_{q,-} &  \mbox{for $q<0$}
                             \end{array}
                             \right. \\*
\hat{b}_{q,s}  \equiv  \left(\frac{\pi}{|q|L} \right)^{1/2} \times
                             \left\{ \begin{array}{ll}
                             \hat{\sigma}_{q,+}  & \mbox{for $q>0$} \\*
                             \hat{\sigma}_{q,-} &  \mbox{for $q<0$}
                             \end{array}
                             \right.
\end{eqnarray*}}
\hfill
\parbox{2.0cm}
{\begin{equation} \label{eqn48} \end{equation}}

\noindent These operators describe independent boson degrees of freedom. Again
the kinetic energy can be expressed in terms of the boson operators
and particle number operators
\begin{equation}
\label{eqn49}
\hat{T} = v_{F} \sum_{q \neq 0} |q| \left( \hat{b}_{q,c}^{\dagger}
\hat{b}_{q,c}^{} +\hat{b}_{q,s}^{\dagger} \hat{b}_{q,s}^{} \right)
+c(\hat{N}) \; .
\end{equation}
For the spin independent interaction (\ref{eqn1}) the spin degrees
are {\it not} renormalized by including the interaction, i.e.
$\tilde{v}_{F,s}(q) \equiv v_{F}$. For the charge degrees of freedom
the problem to find the exact eigenstates is equivalent to the
spin-independent problem. The additional factor $\sqrt{2}$ in Eq.
(\ref{eqn48}) modifies Eq. (\ref{eqn9}) to
\begin{equation}
\label{eqn50}
\tilde{v}_{F,c}(q)= v_{F} \sqrt{\left[1+g_{4}(q)/(\pi v_{F}) \right]^{2}
-\left[ g_{2}(q)/(\pi v_{f})\right]^{2}} \; .
\end{equation}
For the propagators the changes are more dramatic. If one denotes the
propagator for the spinless model by
$G_{\alpha}^{ \mbox{\tiny ${ > \atop (<) }$ } }(x,t;g_{2},g_{4})$ one
obtains using the bosonization of the fermion field operators
\begin{equation}
\label{eqn51}
i G_{\alpha,\sigma}^{ \mbox{\tiny ${ > \atop (<) }$ } }(x,t;g_{2},g_{4})
=\left(i G_{\alpha}^{ \mbox{\tiny ${ > \atop (<) }$ } }(x,t;0,0)
i G_{\alpha}^{ \mbox{\tiny ${ > \atop (<) }$ } }(x,t;2g_{2},2g_{4})
\right)^{1/2} \; .
\end{equation}

For the $g_{4}$-model the more intuitive approach to calculate the
spectral functions via the Lehmann representation described for the
spinless model in Sec. \ref{sec3} can be generalized and leads to a
combinatorical problem slightly more complicated than in Eq.
(\ref{eqn21}).

In the following we restrict the discussion to the step model
(\ref{eqn22}). For the $g_{4}$-model the spectral function has the
same form as (\ref{eqn32})
\begin{equation}
\label{eqn52}
\rho^{>}(k_{n},\omega)=\sum_{l=0}^{n} \tilde{\alpha}_{n-l}^{(n_{c})}
\beta_{l}^{(n_{c})} \delta \left(\omega-\frac{2\pi}{L}  \left[
n  v_{F} +l \left( \tilde{v}_{F}-v_{F} \right) \right] \right) \; .
\end{equation}
The coefficients $\beta_{l}^{(n_{c})}$ are obtained from the power
series with the coefficients $b_{m}^{(n_{c})}(2g_{4})$ by taking the
square root, i.e.
\begin{equation}
\label{eqn53}
\beta_{l}^{(n_{c})} = \left(b_{l}^{(n_{c})} -\sum_{j=1}^{l-1}
\beta_{l-j}^{(n_{c})}  \beta_{j}^{(n_{c})} \right) /\left( 2
\beta_{0}^{(n_{c})} \right) \; .
\end{equation}
The coefficients $\tilde{\alpha}_{n-l}^{(n_{c})}$ are obtained from
$\tilde{a}_{m}^{(n_{c})} = \sum_{n=0}^{m} a_{n}^{(n_{c})}(2g_{4})$ by
the same procedure. For the $g_{4}$-model {\it including} spin the
spectral density differs from a simple delta peak already for
$k<k_{c}$. For $l \leq n_{c}$ the coefficients
$\tilde{\alpha}_{l}^{(n_{c})}$ and $\beta_{l}^{(n_{c})}$ can be given
explicitely as $\tilde{a}_{m}^{(n_{c})}=1=b_{m}^{(n_{c})}$ for $m
\leq n_{c}$. For $1 \ll l \ll n_{c}$ one has $\tilde{\alpha}_{l}^{(n_{c})}
=\beta_{l}^{(n_{c})} \sim l^{-1/2}$. In the limit $L \rightarrow
\infty$ this yields for $0<k<k_{c}$ and $\tilde{v}_{F} \equiv
\tilde{v}_{F,c}>v_{F}$
\begin{equation}
\label{eqn54}
\rho^{>}(k,\omega)= \mbox{const} \, \times \left(\left[\omega-
v_{F}k\right] \left[ \tilde{v}_{F}k-\omega \right] \right)^{-1/2}
\Theta( \omega -v_{F}k) \Theta( \tilde{v}_{F}k-\omega )
\end{equation}
as discussed previously \cite{11,16,17}. Note that this non-Fermi
liquid like behaviour occurs, but one still has $n(k)=0$ for
$k>k_{F}=0$. The behaviour of the spectral function is shown in Fig.
5 (dotted curves). Even for $k \gg k_{c}$ there remains a square root
singularity at threshold which is ``intermediate'' between the delta
peak
expected from the first factor in Eq. (\ref{eqn51}) and the plateau
from the second factor.

For the general model including spin the spectral function follows
from Eq. (\ref{eqn51}) using Eq. (\ref{eqn40})
\begin{eqnarray}
\label{eqn55}
\rho^{>}(k_{n},\omega) & = & A^{-s^{2}}  \sum_{r=0}^{\infty}
\sum_{j=0}^{n+r}  \tilde{\alpha}_{n+r-j}^{(n_{c})}
\beta_{j}^{(n_{c})}
\gamma_{r}^{(n_{c})}
 \nonumber \\*
&& \times \delta \left(\omega-\frac{2\pi}{L}  \left[ \left(
n+r-j \right) v_{F} +\left( r+j \right) \tilde{v}_{F} \right] \right)
\end{eqnarray}
where the coefficients $\beta_{j}^{(n_{c})}$ and $\gamma_{r}^{(n_{c})}$
follow from the coefficients $b_{m}^{(n_{c})}(2g_{2},2g_{4})$ and
$c_{l}^{(n_{c})}(2g_{2},2g_{4})$ defined in Eq. (\ref{eqn41}) by the
procedure (\ref{eqn53}) to take the square root of a power series,
while the $\tilde{\alpha}_{i}^{(n_{c})}$ follow from the
$\tilde{a}_{m}^{(n_{c})}=\sum_{i=1}^{m}
a_{i}^{(n_{c})}(2g_{2},2g_{4})$ in the analogous way.
In the low energy regime Eq. (\ref{eqn55}) leads to the power law
behaviour discussed in detail in references \cite{11} and \cite{17}.
This and the spectral behaviour for larger values of $k$ is shown in
Fig. 5 (solid curves), where the results of the full model are
compared to the $g_{4}$-model. For the value of $s^{2}$ used the
nouniversal features of the spectra are rather similiar, i.e. they
can be largely understood examining the much simpler $g_{4}$-model.

Results for the integrated spectral density $\rho^{>}(\omega)$ are
qualitatively the same as the corresponding curves for the spinless
case shown in Fig. 4.

\section{Summary}
\label{sec6}
In the preceeding sections we have presented results for the spectral
functions $\rho_{+}^{>}(k,\omega)$ and $\rho_{+}^{>}(\omega)$
relevant for inverse photoemission. The corresponding functions to
describe photoemission are given by the mirror images of the curves
presented as $\rho_{+}^{>}(k_{F}+\tilde{k},\omega)=
\rho_{+}^{<}(k_{F}-\tilde{k},-\omega)$. In the Luttinger model with
its linear energy dispersion the value of the Fermi momentum is
irrelevant and has been put to zero in Sec. \ref{sec3}-\ref{sec5}. If
we want to describe nonrelativistic electrons this is no longer the
case as $k_{F}$ is proportional to the electron density. In order to
apply results using the Luttinger model one has to keep in mind that
the linearization procedure is only allowed for sufficiently {\it
long range interactions}, i.e. $k_{c}\ll k_{F}$.

In Ref. \cite{3} angular integrated high resolution photoemission
data of quasi one-dimensional conductors are presented which show a
depletion near the Fermi energy and a rather broad peak about one eV
below the Fermi level. As these spectra are related to
$\rho^{<}(\omega)$ the peak below the Fermi level reminds one of the
peak in one of the integrated spectral functions in Fig. 4. But this
spectrum corresponds to an {\it attractive interaction}. Such a peak
was first discussed by Suzumura \cite{13}, who obtained it for the
case $g_{2}>0$, $g_{4}=0$, i.e. a {\it repulsive} $g_{2}$-interaction
but {\it neglecting} the $g_{4}$-interaction. As one can see from Eq.
(\ref{eqn9}) or (\ref{eqn50}) this leads to
$\tilde{v}_{F}(k\rightarrow 0)< v_{F}$, i.e. an {\it increase} of
spectral weight with decreasing $|\omega|$ until the power law factor
$\omega^{\alpha}$ ($\alpha=2s^{2}$ for the spinless model,
$\alpha=s^{2}$ for the model including spin with spin independent
interaction) leads to suppression and the peak emerges. As
discussed in Sec. \ref{sec2} the {\it physical model corresponds to}
$g_{4}=g_{2}$, i.e. it is {\it unphysical} to neglect the
$g_{4}$-interaction for finite range interactions. Therefore the
experimental peak below the Fermi level cannot be explained as a
Luttinger liquid feature for repulsive interactions. In order to explain the
experimental
depletion near the Fermi level a suprisingly large value of the
exponent has to be assumed \cite{14}.

In this paper we have presented the first detailed study of the
nonuniversal spectral properties of the Luttinger model and have
shown both analytically and numerically that a suprisingly rich
variety of spectral features can emerge.

\acknowledgments
The authors would like to thank J. Voit for bringing Ref. \cite{13}
to their attention.

\appendix{}
\label{apena}
In this appendix we calculate the moments
\begin{eqnarray}
\label{eqna1}
\mu_{n}^{>}(k) & \equiv & \int \omega^{n} \rho^{>}(k,\omega) d\omega
\nonumber \\*
& = & i^{n+1} \int \left(\frac{d^{n} G^{>}(x,t)}{dt^{n}} \right)_{t=0}
e^{-ikx} dx
\end{eqnarray}
of the spectral function $\rho^{>}(k,\omega)$ for $n=0,1$ and $2$. As
$iG^{>}(x,t)=\exp{[F(x,t)]}/L$ we have to calculate the derivatives
of $F(x,t)$

\parbox{11.0cm}
{\begin{eqnarray*}
\dot{G}^{>}(x,0) & = & \dot{F}(x,0) G^{>}(x,0)  \\*
\ddot{G}^{>}(x,0) & = & \left( \ddot{F}(x,0)+ \dot{F}^{2}(x,0) \right)
G^{>}(x,0) \; .
\end{eqnarray*}}
\hfill
\parbox{2.0cm}
{\begin{equation} \label{eqna2} \end{equation}}

\noindent The calculations are straightforward only for the
$g_{4}$-model. In the spinless case the function $F(x,t)$ takes the
form
\begin{equation}
\label{eqna3}
F(x,t)=\sum_{n \geq 1} \frac{1}{n}\exp{\left( i \frac{2\pi}{L} n%
\left[ x-\tilde{v}_{F}(n) t\right] \right) }  \; .
\end{equation}
Therefore $F(x,0)=\ln{\left(1-\exp{\left[i \frac{2\pi}{L}
\left(x+i0\right) \right]} \right)  }$, i.e. $G^{>}(x,0)$ is
identical to the noninteracting Green's function
\begin{equation}
\label{eqna4}
iG^{>}(x,0)=\frac{1}{L} \sum_{n=0}^{\infty}
\exp{\left(i\frac{2\pi}{L} n \left[x+i0\right] \right)} \; .
\end{equation}
The spatial Fourier transform in (\ref{eqna1}) can easily be performed using
(\ref{eqna2}-\ref{eqna4}) and yields
($\tilde{v}_{F}(i)=v_{F}+g_{4}(k_{i})/(2\pi)$)
\begin{eqnarray}
\label{eqna5}
\mu_{1}^{>}(k_{n}) & = & \frac{2\pi}{L} \sum_{m=1}^{n}
\tilde{v}_{F}(m)  \\   \label{eqna6}
\Delta_{n}^{>} & \equiv & \mu_{2}^{>}(k_{n}) - \left(
\mu_{1}^{>}(k_{n}) \right)^{2} \nonumber \\*
& = & \left( \frac{2\pi}{L}\right)^{2} \frac{1}{2} \sum_{j=1}^{n}
\sum_{i=n-j}^{n} \left(\tilde{v}_{F}(i)- \tilde{v}_{F}(j) \right)^{2}
\; .
\end{eqnarray}
For the step model (\ref{eqn22}) this leads for $n>2n_{c}$ to
\begin{equation}
\label{eqna7}
\Delta_{n}^{>}= \left( \frac{2\pi}{L}\right)^{2}
\frac{n_{c}(n_{c}+1)}{2} \left(\tilde{v}_{F}-v_{F}\right)^{2} \; .
\end{equation}
In the limit $L\rightarrow \infty$ this reduces to Eq. (\ref{eqn26}).

For the  general spinless model already the zeroth moment
$\mu_{0}^{>}(k_{n})=1-n(k_{n})$ is nontrivial. It is discussed in
Appendix \ref{apenb}. As the expressions for the first and second
moment are rather lengthy we do not present them here.

For the model including spin we restrict ourselves to the
$g_{4}$-model. If we again write $iG_{\sigma}^{>}(x,t)=\exp{[F(x,t)]}/L$
the function $F(x,t)$ is given by
\begin{eqnarray}
\label{eqna8}
F(x,t)=\frac{1}{2} \sum_{n \geq 1} \frac{1}{n} \left( \exp{\left(i
\frac{2\pi}{L}n\left[x-\tilde{v}_{F}(n) t\right] \right) } +
\exp{\left(i\frac{2\pi}{L}n\left[x-v_{F} t\right] \right) } \right)
\; .
\end{eqnarray}
Therefore $G_{\sigma}^{>}(x,0)$  equals the expression on the rhs. of
Eq. (\ref{eqna1}) and the spatial Fourier transform can be performed
as in the spinless case. The first and the second moment are given by
\begin{eqnarray}
\label{eqna9}
\mu_{1,\sigma}^{>}(k_{n}) & = & \frac{2\pi}{L} \sum_{m=1}^{n}
\left( \tilde{v}_{F}(m)+v_{F} \right)/2  \\   \label{eqna10}
\Delta_{n}^{>} & \equiv & \mu_{2,\sigma}^{>}(k_{n}) - \left(
\mu_{1,\sigma}^{>}(k_{n}) \right)^{2} \nonumber \\*
& = & \left( \frac{2\pi}{L}\right)^{2} \frac{1}{2} \sum_{j=1}^{n}
\sum_{i=n-j}^{n} \left(\tilde{v}_{F}(i)- \tilde{v}_{F}(j)
\right)^{2}/4  \nonumber \\*
&& +
 \left( \frac{2\pi}{L}\right)^{2} \sum_{j=1}^{n}
j \left(\tilde{v}_{F}(j)- v_{F}\right)^{2}/4
\; .
\end{eqnarray}
As now $\tilde{v}_{F}(i)=v_{F}+g_{4}(k_{i})/\pi $ the first term on
the rhs. of (\ref{eqna10}) is identical to the result for the spinless model.
For the
step model the additional term is the only contribution to the width
of the spectrum in the interval $0<k<k_{c}$.

\appendix{}
\label{apenb}
In this Appendix we present a short discussion of the momentum
distribution in the ground state for finite systems and in the limit
$L \rightarrow \infty$. For the special interaction $g_{2}(k) \equiv
g_{4}(k)=g \Theta(k_{c}^{2}-k^{2})$ the Green's function $G^{>}(x,0)$
for the spinless model can be written as
\begin{eqnarray}
\label{eqnb1}
iG^{>}(x,0) & = & \frac{1}{L}
\frac{\exp{\left(ik_{F}\left[N+1\right]x\right)}}%
{1-\exp{\left(i\frac{2\pi}{L}\left[x+i0\right]\right)}} \nonumber
\\*
&& \times \exp{\left(s^{2} \sum_{n=1}^{n_{c}}\frac{1}{n}\left[e^{i(2\pi/L) nx}%
+e^{-i(2\pi /L) nx} -2 \right] \right)} \nonumber \\
& = & \frac{A^{-2s^{2}}}{L}
\frac{\exp{\left(ik_{F}\left[N+1\right]x\right)}}%
{1-\exp{\left(i\frac{2\pi}{L}\left[x+i0\right]\right)}}
\left( \sum_{m=0}^{\infty} c_{m}^{(n_{c})} \exp{\left(%
i\frac{2\pi}{L} mx\right)  } \right) \nonumber \\*
&&  \times \left( \sum_{l=0}^{\infty} c_{l}^{(n_{c})} \exp{\left(%
-i\frac{2\pi}{L} lx\right)  } \right)
\end{eqnarray}
with the coefficients $c_{m}^{(n_{c})}$ given in Eq. (\ref{eqn41})
and
\begin{equation}
\label{eqnb2}
A=\exp{\left( \sum_{n=1}^{n_{c}}\frac{1}{n}\right) }
\stackrel{n_{c} \gg 1}{\rightarrow} e^{C}n_{c} \; .
\end{equation}
The momentum distribution $n(k)$ follows from the spatial Fourier
transform of $iG^{>}(x,0)$
\begin{equation}
\label{eqnb3}
1-n\left( \frac{2\pi}{L} \left[ n_{F}+1+\tilde{n} \right] \right)
= A^{-2s^{2}} \sum_{m=0}^{\infty} \sum_{l=0}^{\infty} d_{m}^{(n_{c})}
c_{l}^{(n_{c})} \delta_{l+\tilde{n},m}
\end{equation}
where $d_{m}^{(n_{c})}=\sum_{l=0}^{m}c_{l}^{(n_{c})}$. The asymptotic
behaviour of $d_{m}^{(n_{c})}$ can be obtained from a comparison of
the power series in (\ref{eqnb1}) for $x=0$
\begin{equation}
\label{eqnb4}
d_{m}^{(n_{c})} \stackrel{m \rightarrow \infty}{\rightarrow}
\sum_{l=0}^{\infty} c_{l}^{(n_{c})} = A^{s^{2}} \; .
\end{equation}
Using (\ref{eqnb4}) the large momentum behaviour
$n\left( \frac{2\pi}{L} \left[ n_{F}+1+\tilde{n} \right] \right)
\rightarrow 0$ for $\tilde{n} \rightarrow \infty$  can be read off
Eq. (\ref{eqnb3}). As $1/2-n(k)$ is symmetric with respect to
$(2\pi/L)
(n_{F}+1/2)$ it is sufficient to consider values $\tilde{n}\geq 0$.

In order to obtain the power law behaviour of the momentum
distribution for $1\ll \tilde{n} \ll n_{c}$ it is useful to calculate
the finite differences $\Delta_{\tilde{n}} \equiv
n\left( \frac{2\pi}{L} \left[ n_{F}+\tilde{n} \right] \right)-
n\left( \frac{2\pi}{L} \left[ n_{F}+1+\tilde{n} \right] \right)$.
They are given by
\begin{equation}
\label{eqnb5}
\Delta_{\tilde{n}}=A^{-2s^{2}} \sum_{l=0}^{\infty} c_{l}^{(n_{c})}
c_{l+\tilde{n}}^{(n_{c})} \; .
\end{equation}
The singular contribution can already be obtained by restricting the
summation in (\ref{eqnb5}) to values $\tilde{n}+l \leq n_{c} $. This
simplifies the discussion as an analytical expression for the
$c_{l}^{(n_{c})}$ (\ref{eqn44}) is available. If one uses the
asymptotic form $c_{m}^{(n_{c})} \sim m^{s^{2}-1}$, Eq. (\ref{eqnb5})
involves a summation over $\left(1/\left[l \left(l+\tilde{n}\right)
\right] \right)^{1-s^{2}} $ which in the thermodynamic limit for
$2s^{2}<1$ contains a singular contribution proportional to
$(1/\tilde{n})^{1-2s^{2}}$. We therefore recover the well known
result \cite{7,8}
\begin{equation}
\label{eqnb6}
n(k)-1/2 \sim \mbox{sign} \,(k_{F}-k) \left| k-k_{F} \right|^{2s^{2}}
\; .
\end{equation}
If one wants to determine the exponent of $\partial n/ \partial k$
numerically from Eq. (\ref{eqnb5}) one has to go to very large values
of $n_{c}$.

\figure{Spectral function $\rho^{>}(k,\omega)$ of the spinless
$g_{4}$-model as a function of $\omega/(v_{F} k_{c})$ for
$\tilde{v}_{F}=1.2v_{F}$ and different momenta as indicated in the
figures. The arrows represents delta peaks with weight $z(k)$.}

\figure{Total spectral density $\rho^{>}(\omega)$ of the spinless
$g_{4}$-model as a function of $\omega/(v_{F} k_{c})$. The solid
curve shows the result for the step model with $\tilde{v}_{F}=1.2
v_{F}$, the dotted curve the result for the exponential model with
$g_{4}/(2\pi v_{F})=0.2$ (repulsive interaction) and the dashed curve
the result for the exponential model with $g_{4}/(2\pi v_{F})=-0.2$
(atractive interaction).}

\figure{The same as in Fig. 1 but for the full $g_{2}=g_{4}$ spinless Luttinger
model for $\tilde{v}_{F}=2v_{F}$. The small oscillations in the full
curves are a finite size effect. The dotted curves shows the
continous part of the related spectral function for the spinless
$g_{4}$-model.}

\figure{The same as in Fig. 2 but for the full $g_{2}=g_{4}$
spinless Luttinger model. The solid
curve shows the result for the step model with $\tilde{v}_{F}=2
v_{F}$, the dotted curve the result for the exponential model with
$g/(\pi v_{F})=3$ (repulsive interaction) and the dashed curve
the result for the exponential model with $g/(\pi v_{F})=-0.49$
(atractive interaction).}

\figure{The same as in Fig. 1 but for the model including spin.
The dotted curves present the
results for the $g_{4}$-model including spin and the full curves the
related results for the full $g_{2}=g_{4}$ Luttinger
model including spin. The
small oscillations are again a finite size effect.}

\end{document}